# Super-diffusive sub-picosecond extraction of hot carriers in black phosphorous

Katsumasa Yoshioka[1,*], Taro Wakamura[1], Takuya Okamoto[1], and Norio Kumada[1]

[1]NTT Basic Research Laboratories, NTT Corporation, 3-1 Morinosato-Wakamiya, Atsugi, 243-0198, Japan

*e-mail: katsumasa.yoshioka@ntt.com

**Abstract:**

**Harvesting hot carriers before they lose energy to the lattice is a critical route toward surpassing the conventional thermodynamic limit in optical-to-electrical (O-E) conversion. However, photocurrent from such hot carriers has remained challenging to directly detect because they equilibrate on picosecond timescales, outpacing conventional electronic measurement. Here, by employing terahertz electronics with sub-picosecond temporal resolution, we directly monitor hot-carrier-driven O-E conversion in black phosphorus (BP). Photoexcitation near the metal contact under zero source-drain bias generates an ultrafast photocurrent with a decay time of ~400 fs—orders of magnitude faster than the typical sub-nanosecond energy relaxation in BP—, demonstrating a measured 3 dB bandwidth of 260 GHz with an intrinsic limit of ~600 GHz. Notably, this photocurrent flows via energetic holes toward the contact electrode, regardless of the equilibrium carrier type, revealing a super-diffusive hot-carrier extraction mechanism. Furthermore, we show that the ultrafast hot-carrier contribution can coexist with the much slower cold-carrier contribution based on the photovoltaic effect, demonstrating that hot carriers can be harvested without discarding lower-energy carriers. These findings highlight the potential of sub-picosecond hot-carrier extraction to expand the O-E conversion bandwidth without sacrificing efficiency, bridging fundamental hot-carrier physics with ultrahigh-speed technological applications.**

**Main text:**

Hot-carrier extraction is increasingly viewed as a key strategy to circumvent the Shockley-Queisser (SQ) efficiency limit in photovoltaic devices[1–5] while also enabling ultrafast optical-to-electrical (O-E) conversion for high-speed data communication[6–9]. The underlying challenge lies in the energy loss on a picosecond timescale, which makes hot carriers difficult to capture and utilize in a practical device. To address this, two-dimensional (2D) van der Waals materials have attracted significant attention[5,10–24] due to their strong quantum confinement and the design flexibility afforded by heterostructuring, both of which can enhance hot-carrier effects. Ultrafast optical spectroscopic studies in these materials have revealed key phenomena, such as efficient carrier multiplication[10,11,23], tunable decay rates and pathways[13,20,24], and interlayer hot-carrier generation and transport[12,18,21,22]. However, such optical techniques cannot capture the actual photocurrent, leaving the ultrafast extraction process at metallic electrodes—the critical step for O-E conversion—largely unresolved.

Among 2D semiconductors, black phosphorus (BP) offers a tunable bandgap, anisotropic carrier transport, and high mobility[25,26] (up to ~10,000 cm$^2$ V$^{-1}$s$^{-1}$). These properties have enabled various photodetector (PD) architectures[27–39], including broadband[27], polarization-sensitive[35], plasmonically enhanced[36,38], and waveguide-integrated[34,37,38] designs. In parallel, combining BP with transition metal dichalcogenides (TMDCs) has yielded p-n junction[29,39] and bulk photovoltaic[30,31] PDs. Although ultrafast optical measurements show exceptional photocarrier diffusion[40,41] and multiplication[42] in BP, the fastest electrically measured 3 dB bandwidth remains limited to ~3 GHz[34]. This indicates that the practical exploitation of hot-carrier effects has been hindered, as the detection of hot-carrier extraction from an electrode requires sub-picosecond electrical readout—well beyond the capabilities of conventional electronics.

Here, we integrate BP with an on-chip, laser-triggered photoconductive (PC) switch[8,43–47], enabling direct access to the previously unresolved ultrafast photocurrent extraction step at the electrode. Under zero bias, the photocurrent decays within ~400 fs, implying an intrinsic 3 dB bandwidth of ~600 GHz and highlighting the importance of hot-carrier extraction. In a gate-tunable device architecture, we identify both a super-diffusive hot-carrier[48–52] component that consistently drives energetic holes to the electrode and a slower cold-carrier component whose polarity depends on doping via the photovoltaic (PV) effect. By harvesting both ultrafast hot and slower cold carriers, our strategy may enable femtosecond optoelectronic devices and enhance energy harvesting efficiency.

**Experimental set-up**

We used three samples with different waveguide and gate structures. The thickness of BP varies between 15 and 25 nm and is indicated in each figure caption. The BP crystal orientation was determined using polarization-resolved Raman spectroscopy (Supplementary Section I) and is noted in the respective figures (Figs. 1b, 2b, and 4a). We begin with the most straightforward device architecture: a mechanically exfoliated BP flake on a sapphire substrate connected to a low-temperature-grown GaAs (LT-GaAs) PC switch via a 10-μm-wide Goubau-line waveguide (Fig. 1a). A femtosecond laser (280 fs pulse width, 517 nm center wavelength, 16.8 MHz repetition) is split into pump and probe beams: the pump excites the BP, while the probe triggers the PC switch at adjustable delays, enabling photocurrent waveform measurement in the time-domain with sub-picosecond resolution. The pump is tightly focused to perform scanning photocurrent microscopy. All measurements were performed at room temperature.

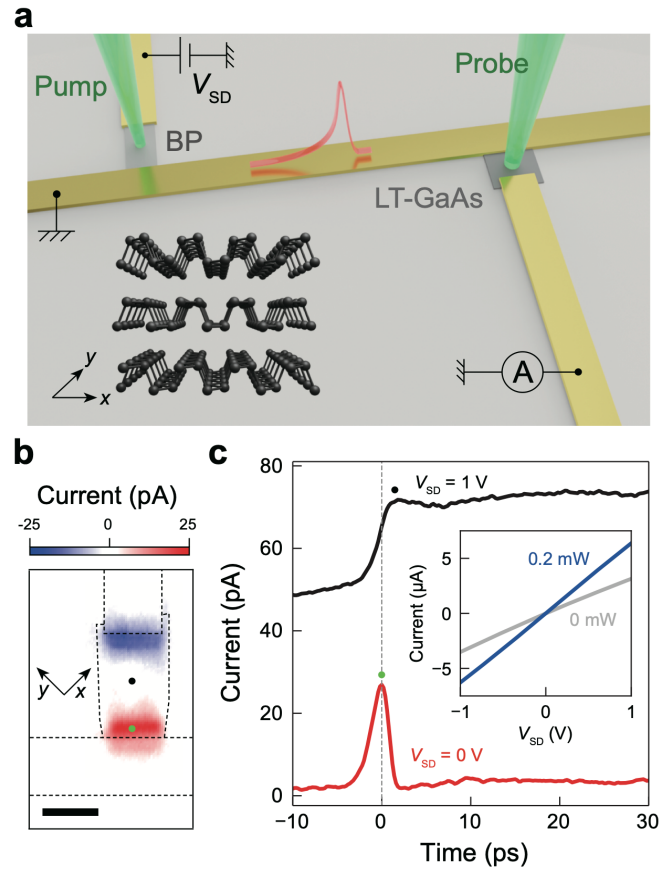

**Figure 1 | Set-up for ultrafast photocurrent readout. a**, Schematic of the device structure. The BP flake is connected to a Goubau line leading to an LT-GaAs PC switch. A pump-probe method captures the photocurrent waveform in the time domain. The inset shows BP's crystal structure with $x$ (armchair) and $y$ (zigzag) axes. **b**, Scanning photocurrent map at $V_{SD} = 0$ V obtained at the peak of transient photocurrent in **c**. Scale bar, 10 μm. **c**, Temporal profiles of the photocurrent with different pump position and bias conditions. Black and green circles in **b** mark the pump spots. The inset shows $I$-$V$ curves with and without the pump beam at the center position (black circle in **b**). BP thickness: 22 nm.

Figure 1b shows a spatial photocurrent map at zero source-drain bias ($V_{SD} = 0$ V) at zero probe beam delay. Photocurrent appears only near the two metallic contacts with opposite signs, similar to previous reports[30,32,53]. Throughout this work, positive photocurrent denotes flow along the waveguide toward the PC switch. Figure 1c compares time-domain photocurrent signals at different excitation spots and $V_{SD}$. As is evident from Fig. 1b, when the center of the flake is excited, no photocurrent is generated at $V_{SD} = 0$. At a finite bias $V_{SD} = 1$ V, a slow photocurrent with no clear decay within 30 ps appears, consistent with the sub-nanosecond energy relaxation reported earlier[40–42,54–57]. The finite offset (~50 pA) before the pump pulse arrives implies incomplete carrier decay between repetitive laser pulses. Increased current flow by photoexcitation confirms a standard PV mechanism, as shown in the inset's I-V data.

In stark contrast, exciting the contact edge (green circle in Fig. 1b) at $V_{SD} = 0$ V generates an ultrafast photocurrent with a full-width at half-maximum (FWHM) of only 2.2 ps, marking the fastest BP PD response to date. Its peak precedes the slower signal by ~2 ps, indicating near-instantaneous extraction from the electrode. We note that observation of this ultrafast response is made possible for the first time by using our approach, which directly measures the photocurrent extracted from the contact with sub-picosecond time resolution. The remainder of this paper explores the physical origin and intrinsic dynamics of this newly discovered ultrafast photocurrent.

**Experimental results**

**Origin of ultrafast zero-bias photocurrent**

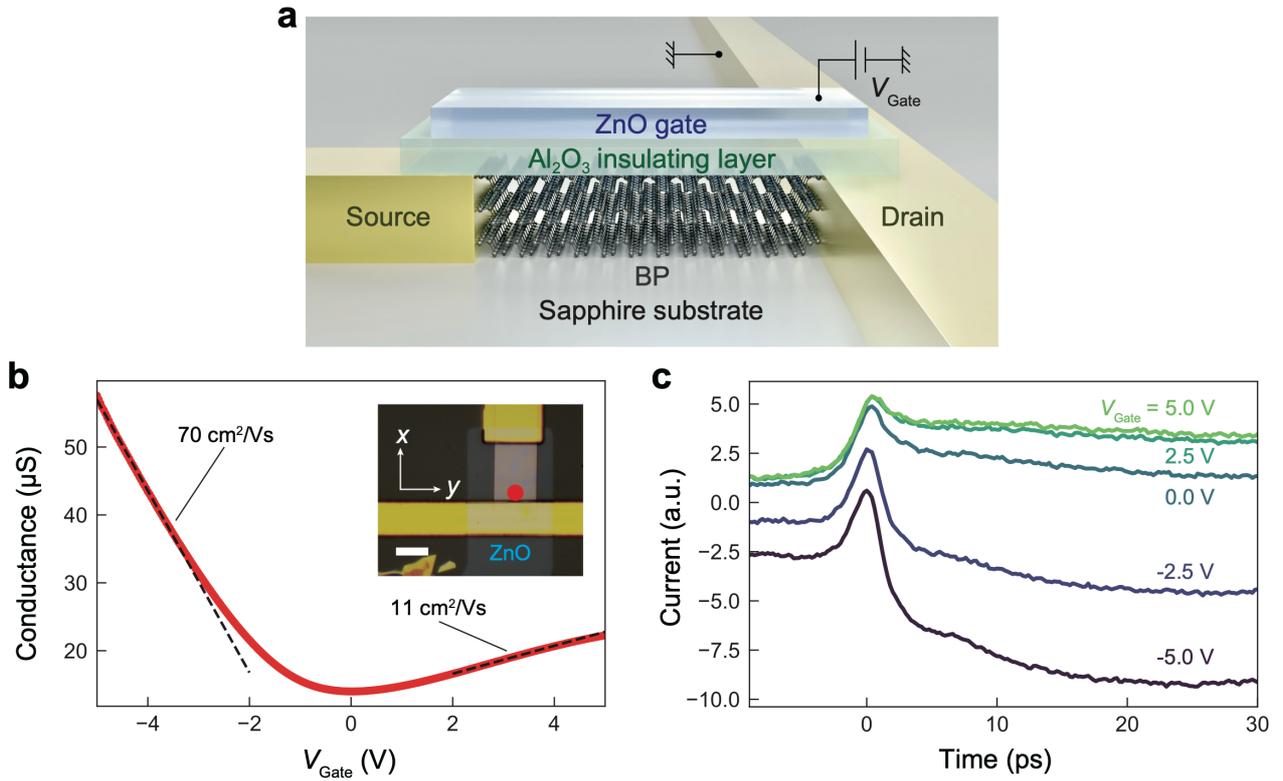

**Figure 2 | Carrier density dependence of photocurrent. a**, Schematic of a BP flake with a ZnO top gate, minimizing the RC time constant and permitting gate control of carrier density. **b**, BP conductance as a function of $V_{Gate}$. Dashed black lines indicate the best linear fits used to determine the field-effect mobility. The inset shows an optical micrograph (scale bar, 10 μm), with the red circle marking the pump spot. **c**, Temporal profiles of the zero-bias photocurrent at the contact edge for different $V_{Gate}$. BP thickness: 25 nm.

To elucidate the origin of the ultrafast photocurrent, we fabricated a top-gated BP device and examined how carrier density affects the measured signals (Fig. 2a). We used zinc oxide (ZnO) as the top gate electrode to minimize the RC time constant from gate capacitance[8,44,58]. Figure 2b plots BP conductance versus gate voltage $V_{Gate}$. The charge neutrality point is located around 0 V, indicating that BP is hole-doped at negative $V_{Gate}$ and electron-doped at positive $V_{Gate}$. The field-effect mobility in the hole-doped regime (70 cm²/Vs) is higher than that in the electron-doped regime (11 cm²/Vs), which agrees with previous studies[25,26,59].

Figure 2c presents the photocurrent at $V_{SD} = 0$ V with the pump beam focused near the contact edge under various $V_{Gate}$ values. An ultrafast pulse as in Fig. 1c appears for all the traces, but a slower component decaying over tens of picoseconds also arises, creating an offset before the pump arrives. This slow component becomes more pronounced at larger doping ($V_{Gate} = \pm 5$ V), which explains the worse visibility in the ungated device. Because these fast and slow signals respond differently to doping, they likely arise from distinct photocurrent generation mechanisms. In BP, multiple decay timescales have been observed by ultrafast optical spectroscopies, spanning tens of femtoseconds[54,60], sub- to a few picoseconds[42,54–57], and extending up to several hundred picoseconds[40–42,54–57]. The fastest response is associated with the thermalization time, reflecting how rapidly a Fermi-Dirac distribution forms through carrier-carrier scattering, whereas the slowest response corresponds to the recombination of cold carriers that exist at the band edge. An intermediate timescale is generally attributed to hot-carrier cooling, where an elevated carrier temperature decreases via carrier-phonon scattering. In our measurements, the observed fast signals fall within the hot-carrier regime, whereas the slow signals correspond to the cold-carrier regime.

To separate these components, we fitted the experimental traces with a sum of a Gaussian (fast) and an exponential rise-decay (slow) function (Supplementary Section II), as illustrated in Fig. 3a. The data are well replicated by this two-term model. Figure 3b shows the fitted amplitudes of each component as a function of $V_{Gate}$. The fast component remains positive for all $V_{Gate}$, indicating a steady flow of photoexcited holes to the contact regardless of whether a BP is hole- or electron-doped. In contrast, the slow response changes sign with $V_{Gate}$: electrons flow to the contact at negative $V_{Gate}$, while holes flow to the contact at positive $V_{Gate}$.

We attribute the consistent ultrafast hole flow to the super-diffusion[48–52] of hot carriers. Immediately after photoexcitation, carriers expand rapidly with a diffusion constant up to ~1000 times higher[48] than that of cold carriers, owing to their elevated effective temperature and kinetic energy. This super-diffusion ends once carriers cool[48–52], matching the picosecond dynamics observed near the contact. Two independent considerations explain why holes dominate the extracted current. First, BP exhibits higher hole mobility than electron mobility[25,26,59], consistent with our gate-dependent transport data in Fig. 2b. Second, ultrafast electron microscopy[41] shows that photoexcited holes diffuse in-plane, whereas electrons diffuse out of plane under the influence of the BP's surface potential. These factors together enhance hole collection efficiency regardless of the equilibrium doping type and distinguish the mechanism from conventional PV or photothermoelectric (PTE) effects. Consequently, energetic holes are consistently extracted at the contact via super-diffusion (Fig. 3d).

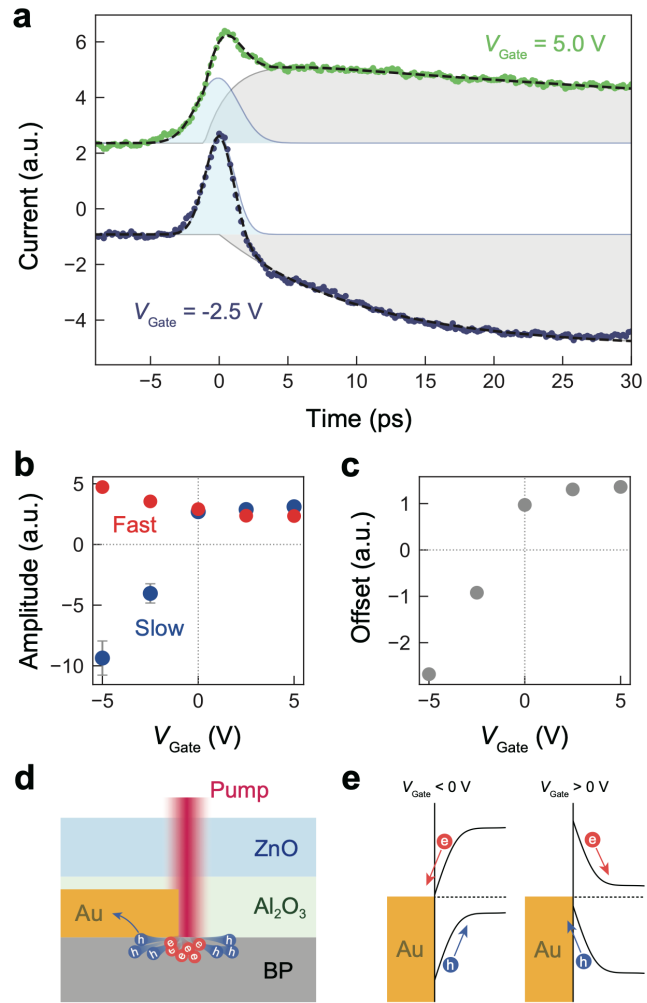

**Figure 3 | Photocurrent generation mechanisms. a**, Experimental (colored) and fitted (dashed black) photocurrent waveforms at $V_{Gate} = -2.5$ V and $+5.0$ V, with Gaussian (fast, light blue) and exponential (slow, gray) components. The trace for $V_{Gate} = +5.0$ V is offset by 1.0 for clarity. **b**, Fitted amplitudes of the fast and slow components as function of $V_{Gate}$. **c**, Fitted offset as a function of $V_{Gate}$. **d**, Schematic of the fast photocurrent origin (super-diffusion effect). **e**, Schematic of the slow photocurrent origin (PV effect under built-in electric field).

On the other hand, the change in the sign of the slow response is consistent with a PV mechanism at the Schottky junction formed at the metal-semiconductor interface[32,53] (Fig. 3e). After the super-diffusion process, the remaining carriers cool to the band edge and are then driven to the electrode by the built-in field, which depends on $V_{Gate}$. The offset before the pump laser pulse (Fig. 3c) follows the same trend as the slow-component amplitude, indicating a long carrier lifetime (>59.5 ns) that rules out a PTE origin for this slow signal.

Both fast and slow photocurrent components in a gate-tunable device need to be controlled to understand photocurrent generation mechanisms in BP. For practical applications, the optimal gate bias can be selected on the basis of the desired outcome: strong doping for maximum energy harvesting, or minimal doping (near the flat band at the Schottky junction) for ultrafast O-E conversion without significant slow response.

**Intrinsic dynamics of hot-carrier extraction**

Next, we investigate intrinsic dynamics of hot-carrier extraction and the effect of crystal anisotropy of BP on it. To this end, we fabricated a device with orthogonal coplanar waveguides (CPWs) aligned to the armchair ($x$) and zigzag ($y$) directions of the BP crystal (Fig. 4a). Compared to the Goubau line, CPWs reduce waveguide dispersion, enabling intrinsic photocurrent dynamics to be more accurately measured. The CPW comprises a 10-μm-wide center conductor flanked by two ground planes, each separated from the center conductor by a 7-μm gap. By selecting which PC switch receives the probe beam, we can capture the photocurrent along either the armchair or zigzag directions.

Figure 4b displays photocurrent waveforms under a 1 V bias applied to the electrode opposite to the readout electrode with the center spot excitation, driving a PV-type response. Along the zigzag direction, the response remains slow with no clear decay within the measurement window, similar to Fig. 1c (black trace). A small peak at ~7 ps arises from multiple reflections between the BP contact and the PC switch, which appears at every measurement configuration in this CPW device. In contrast, the armchair orientation exhibits a faster decay (~30 ps) with negligible offset. This difference aligns with the PV picture, where carriers transit more rapidly along the higher-mobility armchair direction[25,40,41], shortening the observed decay.

Figure 4c highlights zero-bias photocurrent at the contact edge for both orientations. Unlike the biased case, the ultrafast waveforms appear nearly identical in armchair and zigzag directions, indicating carrier transit time has minimal influence on the photocurrent decay. This insensitivity aligns with the super-diffusion picture, in which decay dynamics are governed primarily by energy relaxation of hot carriers in momentum space. The FWHM is ~1.7 ps, shorter than the ~2.2 ps from the Goubau-line device (Fig. 1c, red trace) due to reduced dispersion. Note that it still exceeds the CPW's shortest feasible pulse (~1.2 ps[44]) and thus dispersion effect can be ignored. Consequently, the measured photocurrent $j_{meas}(t)$ is the convolution of the photocurrent at the BP ($j_{gen}(t)$) and the response of the detection PC switch ($j_{det}(-t)$), expressed as $j_{meas}(t) \propto j_{gen}(t) * j_{det}(-t)$, where $t$ denotes the pump-probe delay[45]. By fitting an analytical model that also accounts for the femtosecond laser pulse duration[45] (Supplementary Section III), we extract an intrinsic photocurrent decay constant of 410 ± 30 fs.

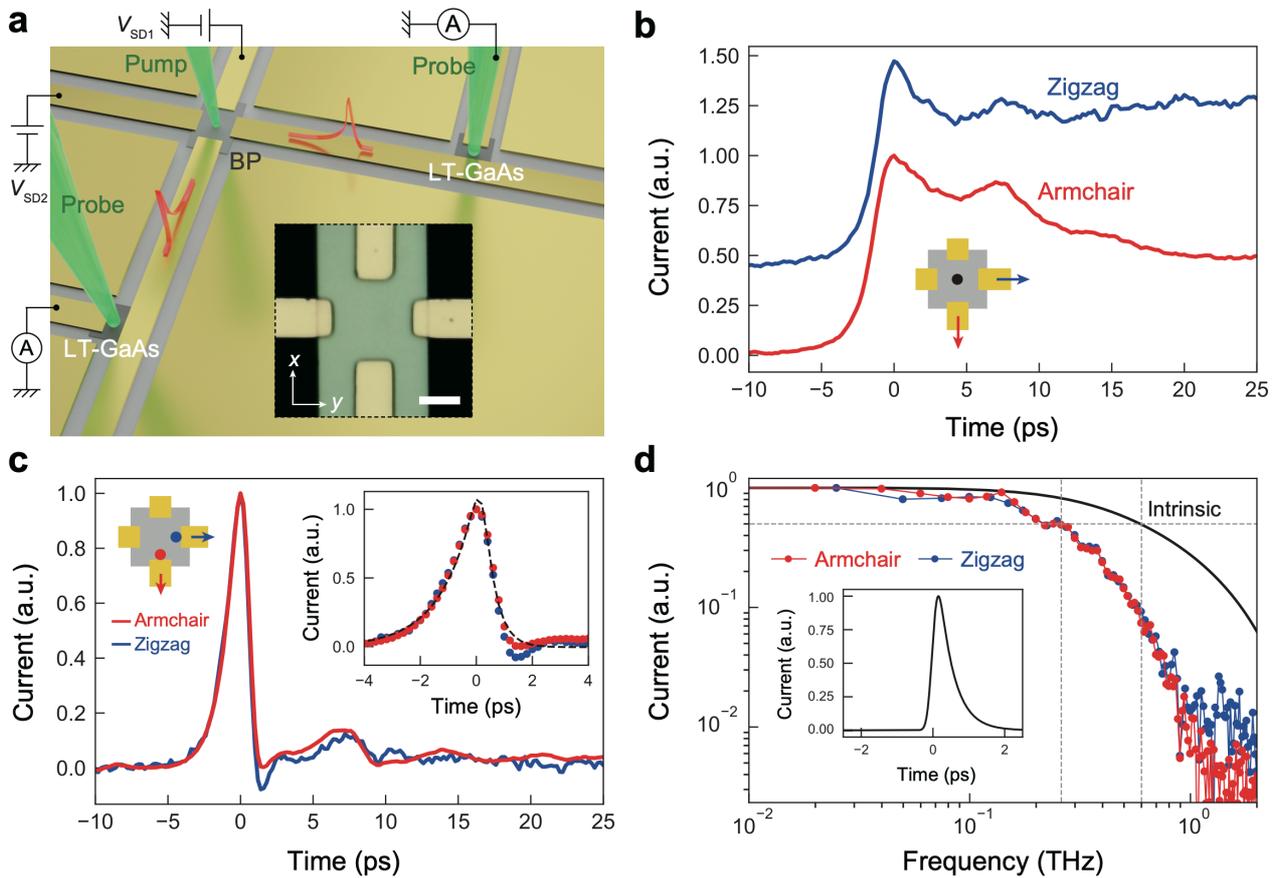

**Figure 4 | Intrinsic dynamics and crystal orientation dependence of photocurrent. a**, Schematic of the device structure. A BP flake is connected orthogonally to two CPWs for measuring armchair ($x$) and zigzag ($y$) directions. The inset shows an optical micrograph (scale bar, 10 μm). **b**, Temporal profiles of the biased photocurrent for zigzag ($V_{SD2}$ = 1 V) and armchair ($V_{SD1}$ = 1 V) directions, normalized to the armchair peak. **c**, Temporal profiles of the zero-bias photocurrent at the contact edge for each orientation, normalized to the peak after subtracting a minor offset. The inset shows a magnified view near the peak position. The dashed curve shows the best fit to the armchair trace obtained using the analytical convolution model. **d**, Fourier-transformed spectra obtained from the waveforms in **c** (colored traces) and estimated intrinsic response (black curve). Vertical dashed lines indicate 3dB bandwidth of 260 and 600 GHz. The inset shows estimated intrinsic photocurrent waveform. BP thickness: 15 nm.

The 3 dB bandwidth of our BP PD estimated by Fourier-transforming the measured time-domain waveform is approximately 260 GHz (Fig. 4d). Furthermore, by removing the bandwidth limit imposed by the detection PC switch, we infer an intrinsic 3 dB bandwidth near 600 GHz. This bandwidth is derived by Fourier-transforming the waveform (inset of Fig. 4d) after removing the convolution effect of $j_{det}(-t)$ (Supplementary Section III). These values are nearly two orders of magnitude higher than the previously reported maximum of ~3 GHz achieved by the PV effect[34], and the inferred intrinsic value exceeds even the fastest graphene PDs reported to date (~500 GHz[9]). Our O-E conversion also outperforms the 68–488 ps intrinsic photocurrent response times derived under identical excitation conditions (contact-edge excitation, zero bias) using a photocurrent autocorrelation technique[31], which utilizes photocurrent saturation effects between two consecutive femtosecond laser pulses. In that approach, the decay time reflects the overall energy relaxation of photoexcited carriers, including recombining cold carriers as well as thermalizing and cooling hot carriers. In contrast, our time-resolved measurements directly isolate the ultrafast photocurrent driven by hot carriers, which precedes and is distinct from any slower cold-carrier effects. This distinction arises because hot carriers are extracted via super-diffusive transport, whereas cold carriers are collected through built-in electric fields. Without direct sub-picosecond readout at the contact, this hot-carrier contribution would remain unresolved.

**Discussion**

In this study, we identified an ultrafast photocurrent in BP with a decay time of approximately 400 fs, a measured 3 dB bandwidth of 260 GHz, and an estimated intrinsic bandwidth of 600 GHz. The remarkable features of this response—including its sub-picosecond lifetime, minimal dependence on transport direction, and consistent flow of photoexcited holes irrespective of doping type—are most convincingly explained by a super-diffusion[48–52] mechanism. Our direct readout of ultrafast photocurrent, which captures non-local hot-carrier transport at the semiconductor-metal interface, proved essential for revealing this process. While conventional optical pump-probe techniques provide valuable insights into energy relaxation near the excitation spot, they do not directly access the spatial transport and extraction of hot carriers at contact interfaces—an essential aspect of ultrafast O-E conversion.

Super-diffusion of photoexcited carriers has been observed in a wide range of materials (including semiconductor[48], metal[49], perovskite[50], polymer[61], and TMDCs[51,52]) by ultrafast microscopy methods. While a transient dipole formed by asymmetric carrier distribution has been used to emit broadband THz radiation via the photo-Dember effect [47,62,63], the microscopic origin and device implications of super-diffusion are distinct and have not yet been exploited for ultrafast hot-carrier extraction in optoelectronic applications. We selected BP for its high carrier mobility, but the super-diffusion-based O-E conversion demonstrated here is likely universal and applicable to a broad class of materials. Notably, we also observed the coexistence of super-diffusion and a conventional PV effect, indicating that extracting hot carriers does not inherently reduce overall O-E conversion efficiency by discarding cold carriers. Instead, both energetic and relaxed carriers can be harvested, effectively broadening the O-E conversion bandwidth and efficiency. From a PD design perspective, one can exploit super-diffusion simply by placing the metallic contact close to the photoexcitation region, enabling carriers to be collected before they cool. While we employed a basic source-drain channel layout to elucidate fundamental aspects of super-diffusion, device structures incorporating waveguide integration or metamaterial engineering could amplify this effect by maximizing the fraction of carriers reaching the contact in their hot state.

We believe that these findings introduce a robust strategy for enhancing PD performance and O-E conversion across various material platforms. By merging super-diffusion with existing device principles, the ultrafast extraction of hot carriers can offer a distinct performance advantage, which will be highly beneficial for developing ultrafast optoelectronics.

**Methods**

**Device fabrication**

PC switches were prepared using a LT-GaAs wafer supplied by BATOP, GmbH. The wafer consisted of a 2.6-μm-thick LT-GaAs surface layer (grown at 300 °C) and a 500-nm-thick $Al_{0.9}Ga_{0.1}As$ sacrificial layer on a semi-insulating GaAs substrate. After patterning the LT-GaAs layer into 100 μm × 100 μm squares with a citric acid solution, the sacrificial layer was dissolved in hydrochloric acid. The resulting LT-GaAs chips were then transferred onto a sapphire substrate using a thermoplastic methacrylate copolymer (Elvacite 2552C, Lucite International) as an adhesive[64]. Residual Elvacite on the sapphire surface was removed with citric acid.

BP was obtained by mechanically exfoliating bulk crystal onto a silica (285 nm)/doped silicon substrates. BP flakes with thicknesses of ~15–25 nm were identified using atomic force microscopy. By using Elvacite, the selected BP flake was picked up and transferred onto the sapphire substrate. Next, Ti/Au waveguides and contacts to the BP were deposited by vacuum evaporation. The entire surface was covered with a 30-nm-thick alumina ($Al_2O_3$) insulating layer grown by atomic layer deposition (ALD). For the ZnO gate device, a 20-nm-thick ZnO top gate (ALD at 140 °C) was patterned on the $Al_2O_3$ layer using photolithography and liftoff. An additional $Al_2O_3$ layer was then deposited to protect the ZnO gate. Finally, to provide electrical access to the waveguide, $Al_2O_3$ on the bonding pads was selectively removed using Miroposit 351 developer.

**On-chip terahertz spectroscopy measurements**

A femtosecond laser (Monaco, Coherent, Ltd) was used as the light source, and the second-harmonic wavelength of 517 nm was generated using a beta barium borate crystal. Two orthogonally polarized pump and probe beams were combined by a polarization beam splitter and aligned with a slight displacement to focus them onto the BP and the LT-GaAs PC switch, respectively, through an objective lens[65]. The pump power was 0.2 mW (horizontal polarization), with a spot size of ~1.8 μm (FWHM). The position of the pump beam was controlled using a motorized mirror, while the probe beam's position was kept constant throughout the experiments. An optical chopper modulated the pump beam at a few hundred hertz for lock-in detection of the terahertz current. All measurements were performed under vacuum to minimize sample degradation and environmental effects.

## Data availability

The datasets generated during and/or analyzed during the current study are available from the corresponding author on reasonable request.

## Acknowledgements

We thank Y. Yamashita and K. Sasaki for fruitful discussions and H. Murofushi for technical support. This work was supported by JSPS KAKENHI Grant Number JP24H00828. This work was conducted as a part of ELEQUANT project, which has received funding from European Innovation Council and SMEs Executive Agency under Grant Agreement 101185712.

## Author contributions

K.Y. and N.K. conceived the experiment. K.Y. designed and built the optical set-up. K.Y. and T.O. performed the measurement. K.Y. analyzed the data. K.Y. and N.K. designed the THz circuits with support from T.W. T. O. and T.W. fabricated the devices. K.Y. and N.K. wrote the paper, with input from all authors.

## Competing financial interests

The authors declare no competing financial interests.

# Sub-picosecond extraction of hot carriers in a black phosphorous photodetector

Katsumasa Yoshioka[1,*], Taro Wakamura[1], Takuya Okamoto[1], and Norio Kumada[1]

[1]NTT Basic Research Laboratories, NTT Corporation, 3-1 Morinosato-Wakamiya, Atsugi, 243-0198, Japan
*e-mail: katsumasa.yoshioka@ntt.com

## I. Polarization-resolved Raman spectroscopy

To determine the crystal orientation of black phosphorus (BP) in the fabricated devices, we performed polarization-resolved Raman spectroscopy (Fig. S1). Figures S1a and S1d correspond to the ungated Goubau-line device (Fig. 1 in the main text), Figures S1b and S1e to the ZnO-gated Goubau-line device (Fig. 2), and Figures S1c and S1f to the CPW device (Fig. 4). A linearly polarized 532 nm laser was focused onto the BP through a 100× objective lens, with polarization controlled by a half-wave plate. The intensities of the $A_g^1$, $A_g^2$, and $B_g^2$ Raman modes were extracted by fitting each spectrum to a Lorentzian function.

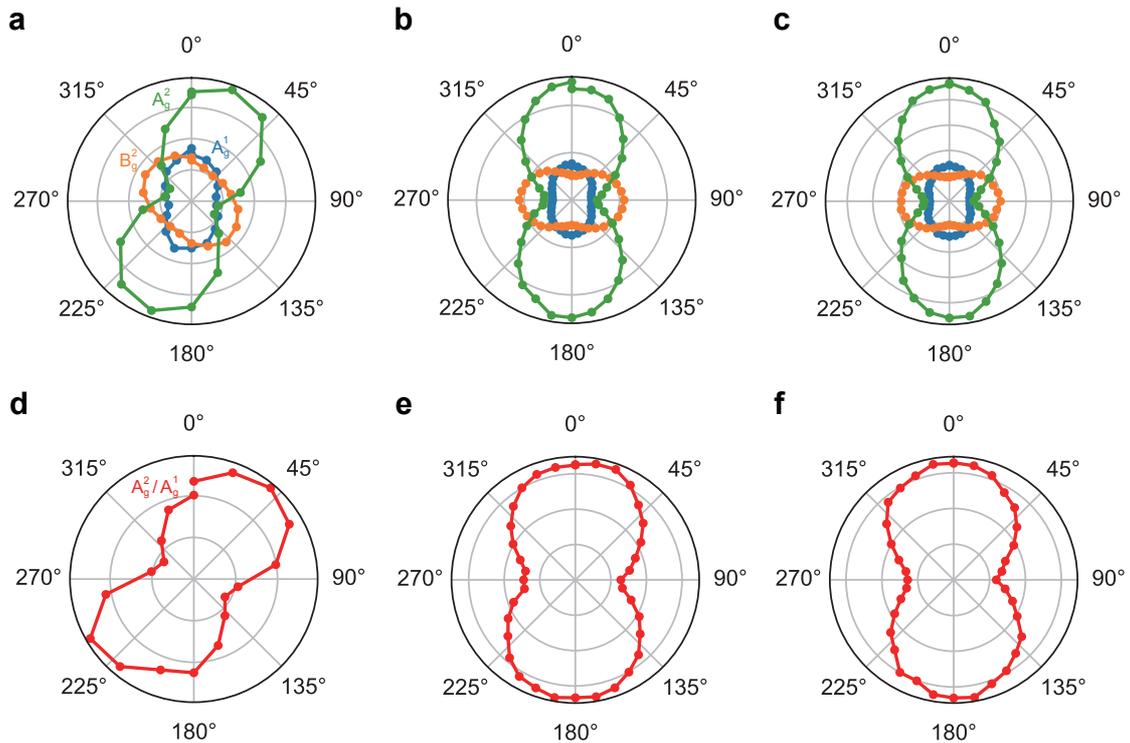

**Figure S1 | Polarization-resolved Raman spectroscopy. a-c**, Intensity of the $A_g^1$, $A_g^2$, and $B_g^2$ Raman modes plotted against polarization angle for the three devices. **d-f**, Raman intensity ratio $A_g^2/A_g^1$, used to identify the armchair ($x$) direction in each device.

We define the armchair direction (*x*) as the angle at which the ratio of the $A_g^2$ to $A_g^1$ Raman intensities ($A_g^2/A_g^1$) is maximized[30,31]. This angle is ~45° for the device in Fig. 1 of the main text and ~0° for the devices in Figs. 2 and 4.

**II. Disentangling fast and slow responses**

To separate the fast and slow components of the gate-tunable photocurrent (Fig. 2c in the main text), we fit the experimental data using a combined Gaussian (fast) and exponential rise–decay (slow) model:

$$A \exp\left(-\frac{t^2}{2\sigma^2}\right) + B\left(\exp\left[-\frac{t}{\tau_{\text{decay}}}\right] - \exp\left[-\frac{t}{\tau_{\text{rise}}}\right]\right) + C,$$

where *A* and *B* are the amplitudes of the fast and slow responses, *C* is the offset, $\sigma^2$ is the variance, and $\tau_{\text{rise}}$ and $\tau_{\text{decay}}$ are the rise and decay time constants, respectively. The resulting fits appear in Fig. S2, while the corresponding fitted values of *A*, *B*, and *C* are shown in Figs. 3b and 3c of the main text.

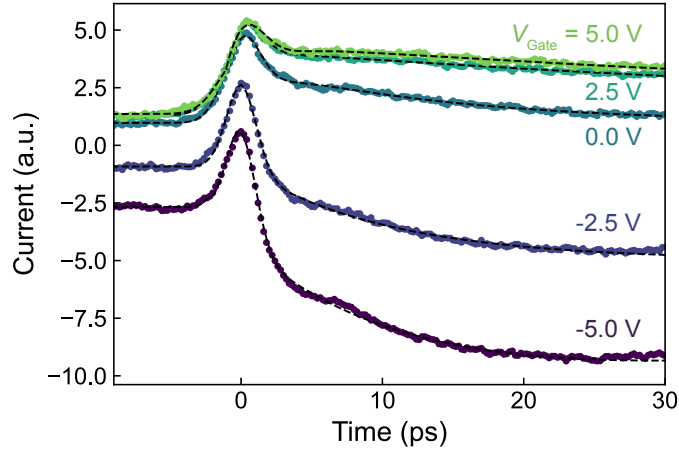

**Figure S2 | Gate-dependent fast and slow photocurrent.** Dashed curves indicate the best fits obtained using the combined Gaussian (fast) and exponential rise–decay (slow) model.

**III. Analysis of photocurrent decay time**

The photocurrent waveform measured in an on-chip geometry can be described as the convolution of generation and detection processes, as developed by P. Zimmermann et al.[45] Specifically,

$$j_{\text{meas}}(t) \propto j_{\text{gen}}(t) * j_{\text{det}}(-t),$$

where $t$ denotes the pump-probe delay. In addition, $j_{\text{meas}}(t)$ itself results from convolving the pump-laser excitation $P_{\text{pump}}(t)$ with the carrier density $n_{\text{gen}}(t)$ in the generation switch (BP):

$$j_{\text{gen}}(t) \propto P_{\text{pump}}(t) * n_{\text{gen}}(t).$$

Here, $P_{\text{pump}}(t)$ is modeled by a Gaussian function as

$$P_{\text{pump}}(t) \propto \exp\left(-\frac{t^2}{2\sigma_1^2}\right),$$

with its standard deviation $\sigma_1$ related to the laser pulse full-width at half-maximum ($t_{p1}$) by $t_{p1} = 2\sqrt{2\ln(2)}\,\sigma_1$. The function $n_{\text{gen}}(t)$ is taken as an exponential decay with carrier lifetime in BP $\tau_1$ multiplied by a unit step function that is zero before the pump arrives:

$$n_{\text{gen}}(t) \propto \exp\left(-\frac{t}{\tau_1}\right) \text{UnitStep}(t).$$

Similarly, the carrier density $n_{\text{det}}(t)$ in the detection switch (GaAs) is modeled as

$$n_{\text{det}}(t) \propto \exp\left(-\frac{t}{\tau_2}\right) \text{UnitStep}(t),$$

where $\tau_2$ is the carrier lifetime in GaAs.

The model assumes a step-like carrier response following excitation, which is reasonable given the short carrier scattering times (tens of femtoseconds) reported in both BP[54,60] and GaAs[66].

Within this framework, the measured photocurrent can be written in analytic form as

$$j_{\text{meas}}(t) = A\left[\exp\left(\frac{\sigma_1^2 + \sigma_2^2 + 2t\tau_1}{2\tau_1^2}\right)\text{erfc}\left(\frac{\sigma_1^2 + \sigma_2^2 + t\tau_1}{\sqrt{2}\,\tau_1\sqrt{\sigma_1^2 + \sigma_2^2}}\right) \right.$$
$$\left. + \exp\left(\frac{\sigma_1^2 + \sigma_2^2 - 2t\tau_2}{2\tau_2^2}\right)\text{erfc}\left(\frac{\sigma_1^2 + \sigma_2^2 - t\tau_2}{\sqrt{2}\,\tau_2\sqrt{\sigma_1^2 + \sigma_2^2}}\right)\right],$$

where $A$ is the amplitude, and the index 1 (2) is the generation (detection) switch. To determine the BP photocurrent decay time, we fit the experimental data (inset of Fig. 4c of the main text) by setting $t_{p1} = t_{p2} = 280$ fs, which is measured via an autocorrelation technique. From this fit, we extract a BP carrier lifetime $\tau_1 = 410 \pm 30$ fs and a detection

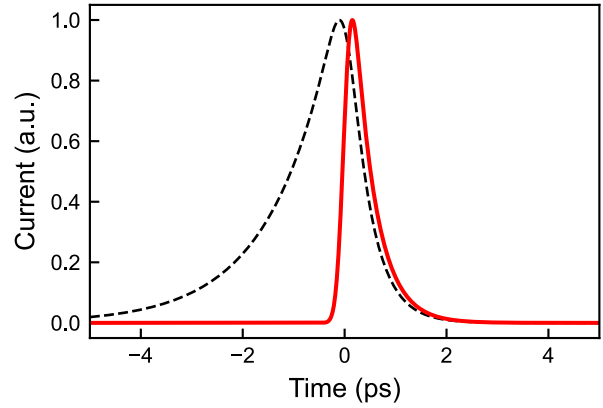

**Figure S3 | Photocurrent waveforms.** The dashed curve shows the best fit to the measured photocurrent using $t_{p1}$ = 280 fs, $\tau_1$ = 410 fs, $t_{p2}$ = 280 fs, and $\tau_2$ = 1210 fs shown in inset of Fig. 4c. The red curve represents the estimated intrinsic photocurrent dynamics obtained by setting $t_{p1}$ = 280 fs, $\tau_1$ = 410 ± 30 fs, $t_{p2}$ = 1 fs, and $\tau_2$ = 1 fs.

PC switch lifetime $\tau_2$ = 1210 fs ± 40 fs.

To estimate the intrinsic photocurrent dynamics at the BP-metal interface (inset of Fig. 4d) by removing the bandwidth limit imposed by the detection PC switch, we set $t_{p1}$ = 280 fs, $\tau_1$ = 410 ± 30 fs, $t_{p2}$ = 1 fs, and $\tau_2$ = 1 fs. The fitted result and the estimated photocurrent waveform are shown in Fig. S3.